\documentstyle[aps,multicol]{revtex}
\begin{document}
\draft
\title{  Scaling Behavior of Driven Interfaces Above the Depinning
Transition  }

\author{Hern\'an A. Makse and Lu\'{\i}s A. Nunes Amaral}

\address{Center for Polymer Studies and Dept. of Physics,
 Boston University, Boston, MA 02215 USA}

\date{\today}

\maketitle

\begin{abstract}

  We study the depinning transition for models representative of each
of the two universality classes of interface roughening with quenched
disorder.  For one of the universality classes, the roughness exponent
changes value at the transition, while the dynamical exponent remains
unchanged.  We also find that the prefactor of the width scales with
the driving force.  We propose several scaling relations connecting
the values of the exponents on both sides of the transition, and
discuss some experimental results in light of these findings.

\end{abstract}


\begin{multicols}{2}

  Interface roughening in the presence of quenched disorder has recently
been the focus of great interest \cite{review}.  In the typical case, a
$d$-dimensional interface described by a height $h({\bf x},t)$ moves in
a $(d+1)$-dimensional disordered medium.  The randomness of the medium
can be described by a quenched disorder $\eta({\bf x},h)$.

  The continual motion of the interface in the medium requires the
application of a driving force $F$.  For driving forces smaller than a
critical value $F_c$, the interface becomes pinned by the disorder after
some finite time.  For $F > F_c$, the interface moves with a constant
velocity $v$.  Hence, the velocity can be identified as the {\it order
parameter\/} of the {\it depinning transition}.  Close to the
transition, the velocity of the interface scales as $v \sim f^{\theta}$,
where $\theta$ is the velocity exponent, and $f \equiv (F-F_c) / F_c$ is
the reduced force.

  Experimental studies of the scaling of the {\it local\/} interface
width, $w \sim \ell^{\alpha}$, where $\ell$ is the window of
observation, reveal roughness exponents $\alpha$ in the range $0.6 -
1.0$ \cite{zhang,exper}.  Several models in which quenched disorder
plays an essential role have been proposed with the goal of explaining
the experimental results (for reviews, see e.g.  \cite{review}).  For
these models, both the scaling of $w$ with $\ell$ and the scaling of the
{\it global\/} interface width $W$ with the system size $L$, were
studied.

  A numerical study \cite{Amaral+Makse}, recently confirmed by
analytical arguments \cite{tkd}, revealed that most of the models
introduced so far could be organized into {\it two\/} universality
classes.  A group of these models can be mapped, at the depinning
transition, onto {\it directed percolation\/} (DP) for $d = 1$, or {\it
directed surfaces\/} (DS) for $d > 1$ \cite{tang,buld}.  These models
are referred to as {\it directed percolation depinning\/} (DPD) models.
The mapping allows the determination of $\alpha$ from the correlation
length exponents of DP or DS.  Reference \cite{Amaral+Makse} showed that
the DPD models can be described by a stochastic differential equation of
the Kardar-Parisi-Zhang (KPZ) type \cite{KPZ86} with quenched disorder
\cite{vicsek}
\begin{equation}
\frac{\partial h}{\partial t} = F + \nabla^2 h+ \lambda
(\nabla h)^2 + \eta({\bf x},h),
\label{qkpz}
\end{equation}
where the coefficient $\lambda$ of the nonlinear term diverges at the
depinning transition.

  A number of different models, belonging to the second universality
class, to which we refer to as {\it quenched Edwards-Wilkinson\/} (QEW),
were found to have either $\lambda = 0$ or $\lambda \rightarrow 0$ at
the depinning transition \cite{Amaral+Makse}.  At the depinning, they
can be described by the EW equation \cite{EW} with quenched disorder
\cite{CDW}
\begin{equation}
\frac{\partial h}{\partial t} = F + \nabla^2 h + \eta({\bf x},h).
\label{qew}
\end{equation}
This equation has been studied by means of the functional
renormalization group \cite{theory}. Numerical studies of the pinned
phase yield $\alpha \simeq 0.97$ \cite{dong} and $\alpha \simeq 1.25$
\cite{lesch1} in $(1+1)$ dimensions.

  A remaining problem is the connection of these universality classes to
the experiments.  We observe that most numerical and analytical studies
focus on the ``pinned phase'' ($F \le F_c$) while nearly all
experimental results are for the ``moving phase'' ($F > F_c$).

  In this Letter, we study the moving phase in $(1+1)$-dimensions for
models representative of each of the two universality classes.  For the
DPD universality class, we find that both $\alpha$ and the growth
exponent $\beta$ --- defined by $W \sim t^{\beta}$ for $t \ll
t_{\times}(L)$, where $t_{\times}(L)$ is the saturation time --- have
different values on both sides of the depinning transition.  However, we
find that the dynamical exponent $z = \alpha / \beta$ has the {\it
same\/} value in both phases.

  First, we consider the QEW universality class.  In
$(1+1)$-dimensions, we study a Hamiltonian model defined as
\begin{equation}
{\cal H} ~ = \sum_{k=1}^L \left [ (h_{k+1} - h_k )^2 - F h_k + \eta(k,
h_k) \right ].
\label{hamilt}
\end{equation}
Here, the first term represents the elastic energy that tends to smooth
the interface, and $\eta(k, h_k)$ is an uncorrelated random number,
which mimics a random potential due to the disorder of the medium.  In
the simulation, a column $k$ is chosen and its height is updated to
$(h_k + 1)$ if the change in (\ref{hamilt}) is negative. Thus, only
motions that decrease the total energy of the system are accepted.
Backward motions are neglected since these are rare events.

  We start by studying the scaling of the local width $w$ in a window of
size $\ell$, for different driving forces.  In Fig. \ref{a+a'}(a), we
show the local width for the pinned and the moving phases.  Consecutive
slopes yield roughness exponents $\alpha_p \simeq 0.92$ for the pinned
phase, and $\alpha_m \simeq 0.92$ for the moving phase (Table
\ref{tab_exp}).  We also detect a dependence of the local width on $L$,
that can be described by a scaling of the form \cite{lesch}
\begin{equation}
w(\ell,L) \sim L^{\alpha_L} \Phi(\ell / L),
\label{self}
\end{equation}
where $\Phi(u)$ is a scaling function that for $u \ll 1$ scales as
$u^{\alpha_p}$.  We find $\alpha_L \simeq 1.23$.

  We study the scaling of $W$ with $t$, and find $\beta_p \simeq
0.85$ and $\beta_m \simeq 0.86$.  These results imply that $z$ remains
unchanged at the transition.

  The study of the scaling of the local width for the moving phase
reveals the existence of two regimes.  In the first
\begin{equation}
w \sim \ell^ {\alpha_m}~f^{-\chi_m} \qquad [\ell \ll \xi],
\label{mov1}
\end{equation}
where $\xi$ is the correlation length, and $\chi_m$ is a new exponent
that characterizes the dependence of the prefactor of the width on the
driving force. In the second regime, the effect of the pinned disorder
becomes irrelevant compared to the annealed disorder \cite{explan}, and
we obtain
\begin{equation}
w \sim \ell^ {\alpha_a}~f^{- \chi_a} \qquad [\ell \gg \xi],
\label{mov2}
\end{equation}
where $\chi_a$ is a new exponent, and $\alpha_a$ is the roughness
exponent corresponding to annealed disorder.  Depending on the absence
or presence of nonlinear terms, we recover the results of either the EW
or the KPZ equation with annealed disorder.  This crossover was observed
both for experiments and simulations of discrete models
\cite{review}.

  Since $\xi \sim f^{-\nu}$, where $\nu$ is the correlation length
exponent, we propose the scaling ansatz
\begin{equation}
w(\ell,f) \sim \ell^{\alpha_a}~ f^{- \chi_a}~ g(\ell / \xi).
\label{scal_w}
\end{equation}
Upon comparison with (\ref{mov1}) and (\ref{mov2}), we find that the
scaling function $g(u)$ satisfies $g(u \gg 1) = const$, and $g(u \ll
1) \sim u^{\alpha_m - \alpha_a}$. We also obtain
\begin{equation}
\chi_a = \chi_m + \nu (\alpha_m - \alpha_a).
\label{chi1}
\end{equation}
In Fig.  \ref{qew_f}, we show the data collapse obtained using the
results of Fig.  \ref{a+a'}(a) according to (\ref{scal_w}).  The
deviation from scaling for large values of $\ell / \xi$ is due to
finite-size effects (see Fig. \ref{dpd_f}(b) for discussion).

  To determine a second relation for the new exponent $\chi_m$, let us
consider the approach of the depinning transition for a system of size
$L$.  For such a system the transition does not occur for $f=0$, as it
would for an infinite system, but for an effective critical force such
that $\xi \sim L$, implying $f \sim L^{-1/ \nu}$.  Replacing this result
into (\ref{mov1}), we obtain
\begin{equation}
w \sim \ell^{\alpha_m}~L^{\chi_m / \nu}  \qquad [\ell \ll L].
\label{mov}
\end{equation}
For the pinned phase we have, according to (\ref{self}),
\begin{equation}
w \sim \ell^{\alpha_p}~L^{\alpha_L - \alpha_p}
\qquad [\ell \ll L].
\label{pin}
\end{equation}
At the transition, (\ref{mov}) and (\ref{pin}) must be identical.  So,
we find $\alpha_p = \alpha_m$, and
\begin{equation}
\chi_m = \nu ~ (\alpha_L - \alpha_p).
\label{chi2}
\end{equation}
Replacing the measured values of $\nu$, $\alpha_L$, $\alpha_p$,
and $\alpha_a$ into (\ref{chi1}) and (\ref{chi2}), we find $\chi_m
\simeq 0.42$ and $\chi_a \simeq 1.04$, in good agreement with the
values obtained in our simulations (Table \ref{tab_exp}).

  Let us now consider the DPD universality class.  For a description of
the model, we refer to the literature \cite{tang,buld}.  In Fig.
\ref{a+a'}(b), we show the local width for the pinned and the moving
phases.  We find $\alpha_p \simeq 0.63$ for the pinned phase, and
$\alpha_m \simeq 0.75$ for the moving phase (Table \ref{tab_exp}).
While previous studies considered $\alpha_m$ as an effective exponent,
and concluded that no self-affine scaling exists in the moving phase for
$\ell \ll \xi$ \cite{tang,buld}, we find a consistent value for
$\alpha_m$ regardless of how closely we approach the transition. This
leads us to suggest the validity of the self-affine scaling for the
moving phase.  This conclusion is supported by the good data collapse
obtained with (\ref{scal_w}), as shown in Fig. \ref{dpd_f}.

  An argument similar to the one leading to (\ref{chi2}) yields
\begin{equation}
\chi_m = \nu ~ (\alpha_p - \alpha_m).
\label{chi3}
\end{equation}
Replacing the known values of $\nu$, $\alpha_p$, $\alpha_m$, and
$\alpha_a$ into (\ref{chi1}) and (\ref{chi3}), we find $\chi_m \simeq
-0.2$ and $\chi_a \simeq 0.23$.  Although the agreement with the
measured values is not exact, the error bars do not rule out the
validity of (\ref{chi3}).

  For the DPD universality class, we also find that $\beta$ changes
values at the depinning transition (see Table \ref{tab_exp}).  We obtain
for both sides of the transition $\alpha \simeq \beta$, implying that
$z$ remains unchanged at the transition \cite{beta}.  The exponent $z$
characterizes the time scale $t_{\times}$ for the propagation of
correlations in the interface. This time scale is not expected to depend
on the external force \cite{havlin}.  These results are in good
agreement with a numerical integration of Eq. (\ref{qkpz})
\cite{vicsek}.

  Next, we compare our numerical results to earlier simulations and
discuss their significance for the determination of the universality
classes of several  experiments.  A problem with the interpretation of
experimental and numerical results for the roughness exponent, has been
the wide range of values for $\alpha$: $0.5-1.0$, measured in the moving
phase.  We note that, in this regime, the crossover to the annealed
disorder regime leads to effective exponents that change with the
velocity (or the driving force).  For this reason, we suggest that the
scaling function (\ref{scal_w}) might be useful in the determination of
the exponents from the study of the local width $w$.  As shown in Table
\ref{tab_exp}, the exponent $\chi_m$  has different signs for the two
universality classes, leading to sharply distinct scaling behaviors for
the prefactor of the width with the driving force.  Since in many
experiments it is possible to monitor the velocity of the interface, and
therefore the driving force, the study of this prefactor may also lead
to an easier identification of the universality class to which the
experimental results belong.

  In light of this discussion, the interpretation of the results of Ref
\cite{kes} is clear.  The numerical integration of the EW equation with
quenched disorder performed in Ref. \cite{kes} must belong to the QEW
universality class, and the reported exponents are effective exponents
whose values were affected by the crossover to the annealed regime.

 The determination of the universality class of the {\it fluid
invasion model\/} (FIM) of Ref. \cite{rob} is not trivial.  However,
we note that $\nu$ is identical for the FIM and for the Hamiltonian
model. Furthermore, the value of $\alpha$ measured in Ref. \cite{rob}
might be explained as an artifact of the use of least square fits to
the local width.

  Since the FIM is a realistic model for the fluid-fluid displacement
experiments of Refs.  \cite{exper}, we propose that they belong to the
universality class of Eq. (\ref{qew}).  In fact, the range of
roughness exponents measured in the experiments of Refs. \cite{exper}
is consistent with the values that could be obtained with the
Hamiltonian model.  Also, in these experiments, it was found that the
global width decreases with the increase of $v$, as for the QEW
universality class.

  The paper burning experiments of Ref. \cite{zhang} and the imbibition
experiments of Ref. \cite{buld} are believed to belong to the DPD
universality class.  This conclusion is supported by the values of the
exponents measured in both experiments.

  In summary, we find that for the DPD universality class, $\alpha$ and
$\beta$ change values at the depinning transition. We also find a new
exponent $\chi_m$, that relates the values of the roughness exponents on
both sides of the transition.  For the QEW universality class, $\alpha$
and $\beta$ remain unchanged, and the exponent $\chi_m$ relates the
different values of the local, $\alpha_p$, and global, $\alpha_L$,
roughness exponents at the depinning transition.

  We thank A.-L. Barab\'asi, S.~V. Buldyrev, R. Cuerno, S. Havlin, S.
Harrington, K.~B.  Lauritsen, P. Rey, R. Sadr-Lahijany, P.-z. Wong, T.
Vicsek and H.~E. Stanley for useful suggestions and discussions.  LANA
acknowledges a fellowship from JNICT.  The Center for Polymer Studies is
supported by NSF.

\begin{figure}
\narrowtext
\caption{  (a) {\it QEW universality class}.  Plot of the local
width $w$ as a function of $\ell$ for several values of the driving
force.  The results were averaged over $30$ realizations of the
disorder and the system size is $5524$.  (b) {\it DPD universality
class}.  Plot of the local width $w$ as a function of $\ell$ for
several values of the driving force.  The system size is $10480$ and
each result was averaged over $50$ realizations of the disorder.  For
clarity, the data for the pinned case is shifted by a factor of $1/2$.
}
\label{a+a'}
\end{figure}

\begin{figure}
\narrowtext
\caption{  {\it QEW universality class}.  Data collapse, according
to (7), of the results displayed in Fig1(a).  }
\label{qew_f}
\end{figure}

\begin{figure}
\narrowtext
\caption{  {\it DPD universality class}.  (a) Data collapse according
to (7), of the results displayed in Fig1(b).  (b) Data collapse for $f =
0.0583$ and systems of different size.  It is visually apparent that the
deviation from scaling observed in (a) is due to finite size effects.  }
\label{dpd_f}
\end{figure}

\end{multicols}

\begin{table}
\narrowtext
\caption{ Critical exponents for the two universality classes studied
in this paper.  The subscript ``$p$'' refers to the pinned phase, the
subscript ``$m$'' denotes the quenched disorder regime in the moving
phase, and the subscript ``$a$'' refers to the annealed disorder regime.
The exponents $z$ and $z_a$ were calculated from scaling relations,
while the remaining exponents were calculated directly in the
simulations. For the QEW, we obtain $\alpha_L = 1.23\pm 0.04$.  }
\begin{tabular}{lcc}
Exponents        & DPD                   & QEW   \\
\tableline
$~\alpha_p$      & $0.63\pm0.03$         & $0.92\pm0.04$ \\
$~\alpha_m$      & $0.75\pm0.04$         & $0.92\pm0.04$ \\
$~\alpha_a$      & $0.50\pm0.04$         & $0.46\pm0.04$ \\
$~z$             & $1.01\pm0.10$         & $1.45\pm0.07$ \\
$~z_a$           & $1.67\pm0.26$         & $2.09\pm0.40$ \\
$~\nu$           & $1.73\pm0.04$         & $1.35\pm0.04$ \\
$~\theta$        & $0.64\pm0.12$         & $0.24\pm0.03$ \\
$~\beta_p$       & $0.67\pm0.05$         & $0.85\pm0.03$ \\
$~\beta_m$       & $0.74\pm0.06$         & $0.86\pm0.03$ \\
$~\beta_a$       & $0.30\pm0.04$         & $0.22\pm0.04$ \\
$~\chi_m$        & $-0.12\pm0.06~~$      & $0.44\pm0.05$ \\
$~\chi_a$        & $0.34\pm0.06$         & $0.99\pm0.05$ \\
\end{tabular}
\label{tab_exp}
\end{table}


\begin{thebibliography}{10}
\bibitem{explan}
Equations (1) and (2) should also include annealed disorder terms.
However, this annealed disorder is much weaker than the quenched
disorder.  For this reason, for $\ell \ll \xi$, the presence of time
dependent disorder is irrelevant.  A different situation occurs for
$\ell \gg \xi$. In this case the effect of quenched disorder is not
relevant, and even a week annealed disorder becomes important.
\bibitem{beta}
Reference \cite{tang} measured $\beta_m \simeq 0.63$. However,
the large intrinsic width \cite{kertesz} of their variant of the
DPD model, leads the authors of Ref. \cite{tang} to add a constant
to their data, which probably altered the scaling of $W$ for
$t \ll t_{\times}$.
\end{thebibliography}
\end{document}